\documentclass{article}

 \usepackage[final]{neurips_2023}

\usepackage[utf8]{inputenc} 
\usepackage[T1]{fontenc}    
\usepackage{hyperref}       
\usepackage{url}            
\usepackage{booktabs}       
\usepackage{amsfonts}       
\usepackage{nicefrac}       
\usepackage{microtype}      
\usepackage[table,xcdraw]{xcolor}         
\usepackage[labelformat = empty]{caption}
\usepackage{multirow}
\usepackage{graphicx}
\usepackage{makecell}
\usepackage[numbers]{natbib}

\definecolor{lightgray}{gray}{0.9}
\definecolor{lightgreen}{RGB}{204, 255, 204}
\definecolor{lightorange}{RGB}{255, 229, 204}
\definecolor{lightred}{RGB}{255, 204, 204}
\definecolor{lightblue}{RGB}{173, 216, 230}
\definecolor{lightpink}{RGB}{255, 182, 193}
\definecolor{darkgreen}{RGB}{0, 100, 0}

\title{Phikon-v2 \\ A large and public feature extractor for biomarker prediction}

\author{%
  Alexandre Filiot$^{*,1}$, Paul Jacob$^{*}$, Alice Mac Kain, Charlie Saillard \\
  Owkin \\
  \texttt{name.surname@owkin.com} \\
  \small{$^{*}$ Contributed equally} \\
  \small{$^{1}$ Corresponding author}
}

\begin{document}

\maketitle

\begin{abstract}

Gathering histopathology slides from over 100 publicly available cohorts, we compile a diverse dataset of 460 million pathology tiles covering more than 30 cancer sites. Using this dataset, we train a large self-supervised vision transformer using DINOv2 \cite{oquab2024DINOv2} and publicly release one iteration of this model for further experimentation, coined \textbf{Phikon-v2}. While trained on publicly available histology slides, Phikon-v2 surpasses our previously released model (Phikon) and performs on par with other histopathology foundation models (FM) trained on proprietary data. Our benchmarks include eight slide-level tasks with results reported on external validation cohorts avoiding any data contamination between pre-training and evaluation datasets. Our downstream training procedure follows a simple yet robust ensembling strategy yielding a +1.75 AUC increase across tasks and models compared to one-shot retraining (p<0.001). We compare Phikon (ViT-B) and Phikon-v2 (ViT-L) against 14 different histology feature extractors, making our evaluation the most comprehensive to date. Our result support evidences that DINOv2 handles joint model and data scaling better than iBOT. Also, we show that recent scaling efforts are overall beneficial to downstream performance in the context of biomarker prediction with GigaPath \cite{prov_gigapath} and H-Optimus-0 \cite{hoptimus0} (two ViT-g with 1.1B parameters each) standing out. However, the statistical margins between the latest top-performing FMs remain mostly non-significant; some even underperform on specific indications or tasks such as MSI prediction - deposed by a 13x smaller model developed internally. While latest foundation models may exhibit limitations for clinical deployment, they nonetheless offer excellent grounds for the development of more specialized and cost-efficient histology encoders fueling AI-guided diagnostic tools.

\end{abstract}

\section{Introduction}
\raggedbottom 
Histopathology is fundamental to disease diagnosis, treatment planning, and medical research. Traditionally, pathologists manually analyze histology slides under microscopes to identify abnormalities, tissue patterns, and disease markers. Over the past years, the advent of digital pathology and the growing availability of Whole Slide Images (WSIs) has facilitated the emergence of computational pathology (CPath), offering great potential for enhancing disease classification, treatment planning and drug discovery through advanced computational methods \cite{baxi_digital_2022, vamathevan_applications_2019}.

Early CPath approaches often relied on transfer learning from ImageNet \cite{deng_imagenet_2009} due to the scarcity of large-scale annotated datasets in digital pathology. Recent advances in self-supervised learning (SSL) have significantly improved representation learning for histopathology, eliminating the need for out-of-domain pre-training. In this paradigm, WSIs are divided into patches and encoded using SSL pre-trained models, providing robust descriptors for various histological features and beyond. These representations are beneficial to the prediction of more complex outcomes such as genomic mutations, survival rates, histological and molecular subtypes \cite{song_artificial_2023}.

With the emergence of improved SSL training methods and the availability of larger digital pathology datasets, several foundation models (FMs) for digital pathology have emerged \cite{prov_gigapath, hoptimus0, vorontsov2024virchow, dippel2024rudolfv, chen2024uni, virchow2}. State-of-the-art models now all leverage DINOv2 \cite{oquab2024DINOv2} method, an extension of iBOT \cite{zhou_ibot_2022} with new pre-training components tailored for data and model scaling. In this work, we introduce Phikon-v2, an improvement over our previous model Phikon \cite{Filiot2023.07.21.23292757}. Phikon-v2 is a Vision Transformer Large (ViT-L) pre-trained with DINOv2 on 460 million histology tiles extracted from more than 55 thousand publicly available slides, covering healthy tissues and more than 30 cancer sites. Through WSI-level biomarker prediction and disease classification tasks, Phikon-v2 demonstrates competitive performance with state-of-the-art models. Our model is publicly available on Hugging Face at \url{https://huggingface.co/owkin/phikon-v2}.

\section{Related Work}
\label{related_work}
\raggedbottom 
Self-supervised learning has allowed researchers to leverage massive amounts of WSIs available through large-scale multicentric datasets such as TCGA\footnote{The Cancer Genome Atlas, available at \url{https://www.cancer.gov/tcga}.}, CPTAC \cite{cptac}, GTEx\footnote{The Genotype-Tissue Expression (GTEx) Portal, available at \url{https://www.gtexportal.org/home/}.} or PAIP\footnote{The Pathology AI Platform, available at \url{http://wisepaip.org/paip}}. Early studies focused on applying SSL methods to pre-train convolutional neural networks (CNN, \cite{cnn_2012}) or ViT \cite{dosovitskiy_vit_2021} architectures on such cohorts to later improve downstream applications. These studies mostly involve patch-level training strategies with various pretext tasks starting by contrastive learning methods \cite{dehaene_self-supervision_2020,saillard_self_2021,schirris_deepsmile_2022,stacke_learning_2022,wang_transpath_2021} such as SimCLR \cite{chen_simclr_simple_2020} or MoCoV2 \cite{chen_moco_2020}, and extending to non-contrastive ones over time \cite{ikezogwo_multi-modal_2022,chen_self_supervised_2022} (for instance, DINO \cite{caron_emerging_2021} or MAE \cite{he_masked_2021}). All of these methods have highlighted the benefits of in-domain SSL pre-training and the use of ViTs over CNNs for a large panel of pre-training methods \cite{kang_benchmarking_2023}.

Over the past years, representation learning in CPath has witnessed significant advances in terms of data and model scaling, in combination with more sophisticated, histology-tailored frameworks. \cite{wang_transformer-based_2022} pre-train CTransPath (a Swin-T architecture of 28M parameters \cite{liu_swin_2021}) on 14.3M patches from TCGA and PAIP, using semantically-relevant contrastive learning, an extension of MoCo v3 \cite{chen_mocov3_2021}. \cite{chen_scaling_2022} introduces HIPT (10M parameters), a hierarchical image pyramid vision transformer pre-trained with DINO on 104M patch-level and 400 thousand region-level images. \cite{kang_benchmarking_2023} publicly released five SSL models (up to 23M parameters) pre-trained on 32.6M TCGA tiles using histology-specific augmentations.

Recent studies have notably marked a turning point for representation learning with the publication of first-ever foundation models in CPath. \cite{azizirobust2023} introduces REMEDIS, a 230M parameter ResNet \cite{he2015deep} pre-trained with SimCLR on 50M pathology tiles from TCGA. \cite{Filiot2023.07.21.23292757} investigates the scaling properties of iBOT \cite{zhou_ibot_2022} and release Phikon, a ViT-Base model  pre-trained on 43M histology tiles from TCGA. More recently, DINOv2 \cite{oquab2024DINOv2} method has been extensively used to develop the latest FMs for histology data. \cite{chen2024uni} proposed and publicly released UNI, a ViT-Large pre-trained on a curated pre-training dataset of over 100M tissue patches from 100 thousand diagnostic H\&E WSIs from GTEx collection and proprietary hospital data. \cite{vorontsov2024virchow} introduces Virchow, a ViT-Huge (632M parameters) pre-trained on 1.5M H\&E slides collected from the Memorial Sloan Kettering Cancer Center. \cite{campanella_computational_2023} pre-train a ViT-S architecture using DINO on the largest pre-training histology dataset made of over 3 billion images extracted from 420 thousand WSIs collected at Mount Sinai Health System. \cite{dippel2024rudolfv} introduces RudolfV, a ViT-L model pre-trained on a curated and diverse dataset of 133 thousand slides (1.2 billion tiles) covering different preparation protocols. Lastly, \cite{kaiko} pre-trained five models from ViT-S to ViT-L using either DINO or DINOv2 methods on 29 thousand WSI (Fresh-Frozen and FFPE) from TCGA. The authors evaluate their model using \texttt{eva}, an open-source framework for evaluating FMs on clinically relevant downstream tasks.

Over the past months, model and pre-training data scaling has reached unprecedented heights starting with Prov-GigaPath \cite{gigapath}, a 1.3 billion parameters model (ViT-g) pre-trained with DINOv2 on 1.3 billion tiles, followed by slide-level pretraining. \cite{hoptimus0} released H-Optimus-0, a ViT-g pre-trained on more than 500,000 WSIs. Lastly, \cite{virchow2} built upon Virchow to propose Virchow2 and Virchow2G, a ViT-H and ViT-G pre-trained using multi-level resolutions on 3.1M WSIs, respectively.

While the current trend of scaling FMs in CPath is a noteworthy step forward, larger models do not systematically guarantee better and more robust representations of histology tissues, especially in the context of biomarker prediction \cite{campanella2024clinicalbenchmarkpublicselfsupervised} which is critical for the clinical deployment of AI solutions \cite{wolflein_good_2023}. \cite{campanella_computational_2023} experimented downstream performance saturation for ViTs pre-trained with more than 325M histology tiles. This observation is consistent with the work of \cite{lai_domain-specific_2023} showing no benefit from training a ViT-B on an uncurated dataset of 600M tiles instead of a ViT-S pre-trained on a curated subset. In the same vein, the results from \cite{alfasly2023rotationagnostic} question the necessity for scaling by showing competitive results on WSI retrieval using only a 9M parameters transformer pre-trained on 6M histology tiles. Lastly, experiments from \cite{kaiko} highlighted no significant benefits of ViT pre-training with DINOv2 over DINO through tile-level evaluation, assuming that the higher complexity of representations learned by DINOv2 can be detrimental to solve simple downstream tasks. Through extensive benchmarks, \cite{campanella2024clinicalbenchmarkpublicselfsupervised} suggests that smaller models perform on par with much larger models ones on most biomarker prediction tasks and are only marginally worse in others.

By comparing Phikon-v2 (ViT-L) to Phikon (ViT-B) and 14 other histology encoders publicly available, we first aim at verifying the potential benefits of DINOv2 over iBOT when scaling together the pre-training dataset size (10x) and model size (4x), then potential performance trends that may arise from FM scaling.

\section{Materials \& Methods}
\label{headings}

\subsection{Training dataset}

\begin{figure}
  \centering
  \includegraphics[width=\textwidth]{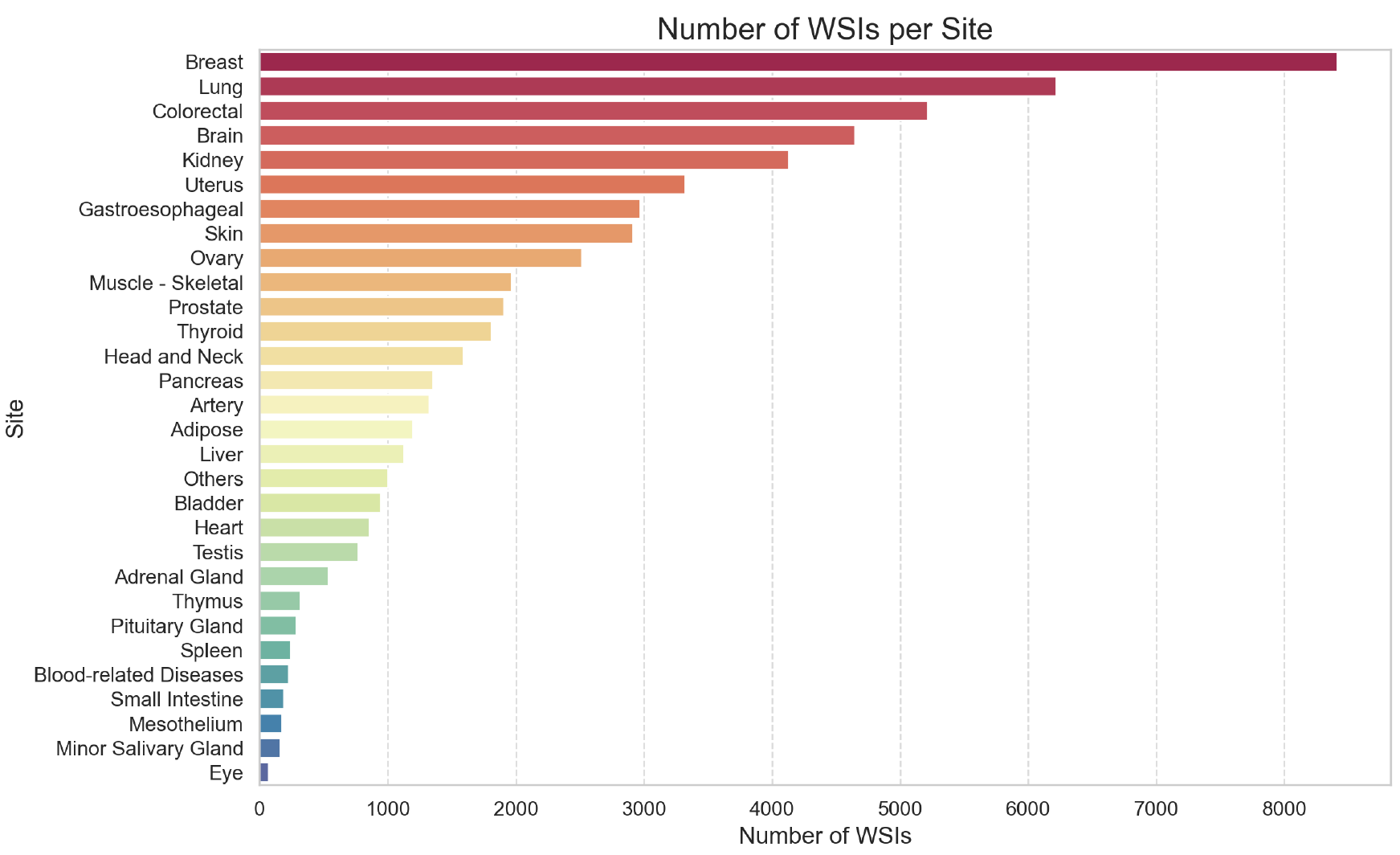}
  \caption{Figure 2: Distribution of tissue sites in PANCAN-XL pre-training dataset. See Extended Table \ref{table:pancanxl_details} for details.}
  \label{slides-per-site}
\end{figure}

\paragraph{PANCAN-XL} Our pre-training set is composed of 132 datasets of digital pathology WSIs publicly available along with 4 internal datasets, covering more than 30 cancer sites and normal tissues for a total of \textbf{58,359 WSIs}. PANCAN-XL is highly heterogeneous with tissue samples (resections and biopsies) collected from various sources, stainings and more than 50 different scanners. The main data sources are CPTAC (6,193 slides) and TCGA (29,502 slides) for malignant tissue, and GTEx (13,302 slides) for normal tissue (23\% of PANCAN-XL). Fresh-Frozen (FF) WSIs accounts for 31\% of the pre-training dataset (17,962 WSIs). Figure \ref{slides-per-site} displays the number of WSIs per site and indication for our pre-training dataset. Of interest, 34\% of PANCAN-XL WSIs come from three indications only (Breast, Lung, Colorectal), while the majority  (approx. 55\%) comes from six different sites (Breast, Lung, Colorectal, Brain, Kidney, Uterus). The exhaustive list of the datasets comprised in PANCAN-XL is available in Extended Table \ref{table:pancanxl_details}.

\paragraph{Preprocessing} Prior to histology tiles extraction, an in-house bi-directional U-Net \cite{ronneberger2015unet}  is
used to segment tissue on the input WSIs and discard
background and artifacts at 2.5x magnification. Using this tissue mask, we then retrieve 224 x 224 histology tiles at magnification 20x (0.5 micrometers per pixel) with a minimal tissue matter proportion of 60\%. At the end of the process, PANCAN-XL is composed of \textbf{456,060,584 tiles} at magnification 20x extracted from the initial 58,359 WSIs.

\subsection{Model and pre-training setup} 

Following current trends in scaling both the dataset and model sizes, we used a vision transformer large (ViT-L) initialized with random weights and trained it on the PANCAN-XL 20x magnification tiles using the DINOv2 self-supervised training method.

Extended Table \ref{table:hyperparams} provides the exhaustive list of hyper-parameters used for pre-training our ViT-L with DINOv2. Most hyper-parameters are taken from the original implementation\footnote{See \url{https://github.com/facebookresearch/DINOv2/blob/main/DINOv2/configs/ssl_default_config.yaml}} with the following adjustment based on iterative findings: our model is trained for 250,000 iterations with a base learning rate of 4e-3 followed by standard cosine schedule. The number of warmup iterations was set to 25,000 and 75,000 for teacher temperature. No registers were used.

Considering that the total number of iterations was set to 250,000 with a batch size of 4,096, this means that roughly 1 billion images were seen during training, which approximately corresponds to twice the size of PANCAN-XL. Note that Phikon-v2 is taken at iteration 100,000, which implies that this model has seen 400M tiles (93\% of PANCAN-XL). Training was performed for 83 hours (11k GPU hours) on 32 nodes with 4 32Go V100 GPUs each, with a total carbon emission of 0.38 tCO2$_{eq}$\footnote{See \url{https://labos1point5.org/les-rapports/estimation-empreinte-calcul} for details.}.

\subsection{Evaluation setup}
\label{evaluation-protocol}

We evaluate and compare different feature extractors on WSI-level weakly-supervised tasks. Those follow a conventional two-step multiple instance learning (MIL) approach. First for each WSI, $N_t$ non-overlapping histology tiles of size 224x224 are extracted at 20x magnification (0.5 micrometers per pixel). These images are passed into a frozen feature extractor yielding a matrix of size $(N_t, d)$, $d$ being the output features dimension. Finally, a permutation-invariant pooling operator is optimized to aggregate patch-level features into a single slide-level prediction. For all tasks, the same exact $N_t=5,000$ tiles were randomly sampled from each WSI across feature extractors (padding is applied for slides with less than $N_t$ tiles with organic tissue). Note that we set $N_t=400$ for ISUP grading task due to memory constraint. Also, we chose the Attention-Based Multiple Instance Learning (ABMIL) algorithm \cite{ilse_attention_based_2018} to perform feature matrices aggregation, which serves as a strong baseline among MIL algorithms \cite{jaume2024multistainpretrainingsliderepresentation}. We use the original one-layer gated architecture with input embeddings mapped to an embedding dimension of 128 and weights attention matrix of dimension $(128, 128)$, without dropout and with sigmoid activation. This makes a total of $164,482$ learnable parameters. ABMIL was trained using the Adam optimizer \cite{kingma2017adam} with parameters (0.9, 0.999), a batch-size of 16 and a constant learning rate of 1e-3. Binary cross-entropy (resp. cross-entropy) was used for binary classification tasks (resp. multi-class classification).

Our training and evaluation protocol is performed as follows:

\begin{enumerate}
    \item For each task, we create 5 case and label-stratified splits on the training dataset that are kept constant across all feature extractors. Then, we perform a 5-fold cross validation by keeping one split apart each time for validation. For each training split, we train an ensemble of 5 models (which differ only from initialization) for 100 epochs (except for the DINOv2 ViT-L ImageNet, for which we use 300 epochs to take into account domain shift). A single cross-validation run thus corresponds to 5x5 trainings, hence 25 different ABMIL models. For each training, the selected model is taken as maximizing AUC on the validation splits among epochs 1, 3, 5, 10, 15, 20, 30, 40, 50, 60, 70, 80, 90 or 100 (complemented by epochs 120, 150, 180, 200, 220, 250, 280 or 300 for DINOv2 ViT-L ImageNet).
    \item Then, for each of the the external validation datasets, we perform inference of the 25 different ABMIL models (with respective optimized epochs) trained during cross-validation. We report the final AUC by using an ensemble of the predictions. We report AUC only on external validation datasets.
    \item Additionally, we compare the previous ensembling approach to one-shot retraining. This standard strategy consists in first retrieving the optimal epoch maximizing the validation AUC averaged over the 25 splits across the epochs 1, 3, 5, 10, 15, 20, 30, 40, 50, 60, 70, 80, 90 or 100. Then, a single ABMIL model is trained on the full training set for exactly the same optimal number of epochs.
\end{enumerate}

Note that, depending on original implementations, some specificities may exist between models:
\begin{itemize}
    \item Virchow and Virchow2 concatenate the \texttt{[CLS]} token and the average pooling of patch tokens (16x16), leading to an embedding dimension $d=2,560$.
    \item Patches of size 512x512 were extracted for PathDINO-512 \cite{alfasly2023rotationagnostic}.
    \item For CTransPath \cite{ctranspath}, tiles were extracted at mpp 1.0 following authors' recommendation.
    \item In view of \cite{jaume2024multistainpretrainingsliderepresentation}, only GigaPath tile-level encoder \cite{gigapath} was used to generate WSI-level feature matrices.

\end{itemize}

\subsection{Evaluation tasks}
\label{evaluation-tasks}

We derive 8 WSI-level tasks coming from 5 different cancer sites. All tasks consider WSIs at the 20x magnification for feature extraction, except for \textbf{RCC (Kidney)} (5x). Note that there is an overlap between TCGA \textit{downstream training cohorts} and our pre-training dataset; however, we make sure to avoid any data leakage between PANCAN-XL and \textit{downstream external validation cohorts} for which we report classification metrics, as this data leakage tends to overestimate generalization error \cite{chen2024uni}.

\textbf{METASTASIS (Breast Lymph Nodes)}: The breast metastasis detection task from the 2016 Cancer Metastases in Lymph Nodes Challenge (CAMELYON16 \cite{ehteshami_bejnordi_diagnostic_2017}) involves 399 H\&E FFPE histopathology whole slide images (WSIs) of sentinel lymph nodes from Radboud University Medical Center and University Medical Center Utrecht. For our training and evaluation process, we utilized the official train-test splits. Train and test sets contain 269 (159 normal, 110 metastasis) and 130 (80 normal, 50 metastasis) WSIs, respectively.

\textbf{MSI (Colorectal)}: MSI
mutation impact the deoxyribonucleic acid damage repair (DDR) process in tumors. Early recognition of this biomarker may benefit the patients through specific therapies. This is of particular interest in colorectal cancer \cite{saillard_blind_2023} where MSI mutation can account for up to 20\%. We aim at predicting high vs low
instability (MSI-H vs MSS/MSI-L). \textbf{Training sets}: We use TCGA-COAD and TCGA-READ as training cohorts, accounting for 611 H\&E FFPE WSIs and 595 patients (511 MSS/MSI-L, 84 MSI-H). \textbf{Evaluation sets}: We use two different external validation cohorts. First one is \textbf{Paip}. The PAIP consortium provides 2,547 H\&E FFPE WSI collected from three Korean centers
(Seoul National University Hospital, Seoul National
University Bundang Hospital and SMG-SNU Boramae
Medical Center), covering six cancer types. We retrieved 47
patients (47 WSIs, 35 MSS/MSI-L and 12 MSI-H) from PAIP with colorectal tumors and available
MSS/MSI-L labels, provided by the Pathology AI Platform. Second cohort is \textbf{Cy1}, which is a private collection of 698 H\&E and H\&E\&S biopsies from 698 patients (450 MSS/MSI-L, 248 MSI-H) digitized in France. For all cohorts, MSI labels were determined using 4-plex IHC and MSI-PCR when available.

\textbf{HER2 (Breast)}: HER2 (encoded by ERBB2 gene) is a protein present in the membranes of cells and controlling their growth. HER2 is ampliﬁed and/or overexpressed in approximately 15-20\% of breast cancers. The overexpression and/or amplification of HER2 has been associated with aggressive clinical behavior but with a high probability of response to HER2 targeted therapy during and/or after chemotherapy, resulting in a significant improvement in disease-free and overall survival. \textbf{Training sets:} We use TCGA-BRCA as training cohort using reliable labels from
\cite{tcga_2012_her2}. This corresponds to 801 H\&E FFPE WSIs (680 HER2-negative, 121 HER2-positive) collected from 752 patients. \textbf{Evaluation sets}: We used two external cohorts. First one is the training set of the \textbf{Herohe} \cite{conde_sousa_herohe_2022} dataset, which contains 360 cases (144 HER2+ and 216 HER2-). Second dataset is \textbf{Bcnb} \cite{bcnb}, a dataset of 1,058 early breast cancer core-needle biopsies (277 HER2+ and 781 HER2-) collected from 1,058 patients along with clinical characteristics.

\textbf{ER (Breast)}: Predicting estrogen receptor (ER) and progesterone receptor (PR) status from WSIs is crucial in breast cancer care. ER and PR are proteins that, when present, indicate that the cancer may respond well to hormonal therapies. \textbf{Training set}: We use TCGA-BRCA as training cohort with labels available from TCGA PanCancer Atlas (2018) \footnote{See https://www.cbioportal.org/study/summary?id=brca\_tcga\_pan\_can\_atlas\_2018.}. Our training set consists of 1,076 H\&E FFPE WSIs (235 ER- and 841 ER+) collected from 1,008 patients.
\textbf{Evaluation sets}: We use the \textbf{Bcnb} dataset containing 1,058 samples (227 ER- and 831 ER+).

\textbf{PR (Breast)}: Similarly to the \textbf{ER (Breast)} task, our training set if composed of 1,073 H\&E FFPE WSIs (344 PR- and 729 PR+) taken from 1,005 patients from \textbf{TCGA-BRCA}. \textbf{Evaluation sets}: We use the \textbf{Bcnb} dataset containing 1,058 samples (268 PR- and 790 PR+).

\textbf{IDH1 (Brain)}: This task is inspired from \cite{chen2024uni}. It consists in predicting one of 5 glioma histomolecular subtypes from glioblastoma, astrocytoma, and oligodendroglioma cases with molecular status. \textbf{Training set}: We use TCGA-GBM and TCGA-LGG for training, which gathers 1,160 H\&E FFPE WSIs from 576 patients with available subtype information. WSI labels are distributed as follows: 154 IDH1-mutant astrocytomas, 66 IDH1-mutant glioblastomas, 232 IDH1-mutant and 1p/19q-codeleted oligodendrogliomas, 78 IDH1-wildtype glioblastomas and 630 IDH1-wildtype astrocytomas. \textbf{Evaluation set}: We use a subset of 873 WSIs (873 patients) from the EBRAINS Digital Tumor Atlas \cite{ebrains} with available IDH1 mutation status. WSI labels are distributed as follows: 123 IDH1-mutant astrocytomas, 34 IDH1-mutant glioblastomas, 176 IDH1-mutant and 1p/19q-codeleted oligodendrogliomas, 66 IDH1-wildtype glioblastomas and 474 IDH1-wildtype astrocytomas.

\textbf{ISUP (Prostate)}: The ISUP grading system for prostate cancer evaluates tumor aggressiveness based on microscopic cell appearance, with grades ranging from 1 to 5. It is crucial for treatment planning as it influences prognosis, treatment decisions, and the personalization of therapy. The ISUP grading task is derived from the Prostate Cancer Grade Assessment challenge (PANDA, \cite{panda}), which comprises 10,616 prostate cancer core needle biopsies of
prostate cancer sourced from the Radboud University Medical Center and the Karolinska Institute. Each slide
is assigned an ISUP score that defines prostate cancer grade (6-class grading task). We used an arbitrary, case and label stratified train-test split. \textbf{Training set}: Contains 9,553 WSIs from 9,553 patients with 2,601 G0, 2,400 G1, 1,209 G2, 1,117 G3, 1,124 G4 and 1,102 G5. \textbf{Evaluation set}: Contains 1,062 WSIs from 1,062 patients with 290 G0, 266 G1, 134 G2, 125 G3, 125 G4 and 122 G5.

\textbf{RCC (Kidney)}: The RCC subtyping task consists in predicting cancer subtypes in cell carcinoma between primary clear cell renal cell carcinoma
(CCRCC), papillary renal cell carcinoma (PRCC) and chromophobe renal cell carcinoma (CHRCC). \textbf{Training sets}: We combined TCGA-KIRC, TCGA-KIRP and TCGA-KICH to form a unique training set comprising 949 H\&E FFPE WSIs collected from 897 patients with the following subtypes distribution: 529 CCRCC, 299 PRCC ad 121 CHRCC. \textbf{Evaluation set}: We use the DHMC cohort \cite{dhmc} which originally comprises 563 H\&E FFPE WSIs of renal cell carcinoma from the Department of Pathology and Laboratory Medicine at Dartmouth-Hitchcock Medical Center. We filtered out oncocytomas resulting in 467 WSIs (344 CCRCC, 100 PRCC and 23 CHRCC).

\subsection{Comparisons \& baselines}

\begin{table}[ht!] 
  \centering
  \addtolength{\leftskip} {-3cm}
  \addtolength{\rightskip}{-3cm}
  \begin{tabular}{llllllllllllll}
    \toprule
    \textbf{Model} & SSL method & \makecell[tl]{Architecture \\ (no. parameters)} &  \makecell[tl]{No. WSIs \\ (no. tiles)} & Origin of WSIs \\
    \midrule

    H-Optimus-0 \cite{hoptimus0} & DINOv2 & ViT-G/14 (1.1B) & 500k (-) & Proprietary  \\
    GigaPath \cite{gigapath} & DINOv2 & ViT-G/14 (1.1B) & 171k (1.4B) & Providence health network$^*$  \\

    Virchow2 \cite{virchow2} & DINOv2 & ViT-H/14 (632M) & 3.1M (-) & MSKCC$^*$  \\
    Virchow \cite{virchow} & DINOv2 & ViT-H/14 (632M) & 1.5M (-) & MSKCC$^*$  \\

    UNI \cite{chen2024uni} & DINOv2 & ViT-L/16 (307M) & 100k (100M) & MGH$^{1,*}$, BWH$^{2,*}$, GTEx  \\
    \rowcolor{lightblue} Phikon-v2 (Ours) & DINOv2 & ViT-L/16 (307M) & 58.4k (456M) & TCGA, CPTAC, GTEx [...] \\

    Remedis \cite{azizirobust2023} & SimCLR & ResNet-152$\times$2 (232M) & 29.0k (50M) & TCGA \\

    Hibou (B/8) \cite{hibou} & DINOv2 & ViT-B/8 (86M) & 1.1M (512M) & Proprietary   \\
    Kaiko (B/8) \cite{kaiko} & DINO & ViT-B/8 (86M) & 29k (256M)  & TCGA \\
    \rowcolor{lightpink} Phikon \cite{Filiot2023.07.21.23292757} & iBOT & ViT-B/16 (86M) & 6.0k (43M) & TCGA \\

    CTransPath \cite{ctranspath} & MoCo-v3 & Swin-T/14  (28M) & 32.2k (16M) & TCGA, PAIP \\
    Lunit-DINO (B/8) \cite{lunit_dino} & DINO & ViT-S/8 (22M) & 36.7k (33M) & TCGA, TULIP$^*$ \\
    PathDINO-512 \cite{alfasly2023rotationagnostic} & DINO & ViT/16 (9M) & 11.8k (6M) & TCGA \\
    
    \midrule
    CONCH \cite{CONCH} & iBOT $\rightarrow$ CoCa \cite{yu_coca_2022} & ViT-B/16 (86M) & 21.4k (16M) $\rightarrow$ 1.2M pairs & TCGA $\rightarrow$ EDU$^3$, PMC OA$^4$   \\
    PLIP \cite{huang2023visual} & CLIP \cite{radford_learning_2021} & ViT-B/32 (86M) & 208k pairs & Twitter, PathLAION \cite{huang2023visual} \\
    \midrule
    DINOv2 ViT-L ImageNet \cite{oquab2024DINOv2} & DINOv2 & ViT-L/14 (307M) & 142M natural images & LVD-142M \\
    \bottomrule
\end{tabular}
\vspace{0.5cm}
\caption{Table 1: Presentation of the different histology feature extractors evaluated in our benchmark. Models are sorted by decreasing size and number of WSIs. CONCH \cite{CONCH} and PLIP \cite{huang2023visual} are pre-trained using image-text pairs. iBOT $\rightarrow$ CoCa indicates that the vision backbone was first pre-trained with iBOT then fine-tuned with image-text pairs using CoCa contrastive method. Last raw corresponds to the baseline pre-trained on natural images. $^*$ indicates private datasets. $^1$: Massachusetts General Hospital, $^2$: Brigham \& Women’s Hospital, $^3$: Educational sources, $^4$: PubMed Central Open Access Dataset.}
\label{table:feature-extractors}
\end{table}

We compare Phikon-v2 against 14 of the most recent feature extractors publicly available to date. Table \ref{table:feature-extractors} summarizes each of the corresponding models in terms of architecture, self-supervised pre-training scheme and dataset. Our baseline is a ViT-L/14 distilled from a ViT-G/14 pre-trained with DINOv2 on LVD-142M \cite{oquab2024DINOv2}, a curated dataset of 142 million natural images, which we coin "DINOv2 ViT-L ImageNet" for clarity.

\subsection{Statistical analysis}

AUC is reported as the area under the receiver operating curve plotting true positive rate against the false positive
rate as the classification threshold is varied. 95\% confidence intervals are computed using bootstrapping with 10,000 repeats unless specified otherwise. AUCs between two models are compared using permutation tests \cite{comp_auc} with Holm's correction \cite{holm1979simple}. Combined p-values are computed using Fisher's method \cite{fisher1925statistical}.

\section{Results}
\label{results}

\begin{table}[ht!] 
  \centering
  \addtolength{\leftskip} {-3cm}
  \addtolength{\rightskip}{-3cm}
  \begin{tabular}{lllllllllll|l}
    \toprule
    \multicolumn{1}{c}{} & 
    \multicolumn{1}{c}{ER} & 
    \multicolumn{1}{c}{PR} & 
    \multicolumn{2}{c}{HER2} &
    \multicolumn{1}{c}{IDH1} &
    \multicolumn{1}{c}{ISUP} &
    \multicolumn{1}{c}{META.} &
    \multicolumn{2}{c}{MSI} &
    \multicolumn{1}{c}{RCC} \\
    \cmidrule(r){2-2}
    \cmidrule(r){3-3}
    \cmidrule(r){4-5}
    \cmidrule(r){6-6}
    \cmidrule(r){7-7}
    \cmidrule(r){8-8}
    \cmidrule(r){9-10}
    \cmidrule(r){11-11}
    \textbf{Extractor} & Bcnb & Bcnb & Bcnb & Herohe & Ebrains & Panda & Cam16 & Cy1 & Paip  & Dhmc & \textbf{Average} \\
    \midrule
    GigaPath \cite{gigapath}  & \textbf{0.878} & \textbf{0.836} & \textbf{0.736} & 0.723 & \textbf{0.895} & \textbf{0.944} & 0.995 & \textbf{0.888} & 0.980 & \textbf{0.996} & \textbf{0.883}$^{\textcolor{darkgreen}{0.020{{\uparrow}}}}$ \\
    
    \textbf{Phikon-v2 (Ours)} & 0.856 & 0.804 & 0.669 & \textbf{0.770} & 0.842 & 0.936 & 0.997 & \underline{0.882} & \textbf{0.991} & 0.989 & \underline{0.874}$^{\textcolor{darkgreen}{0.011{{\uparrow}}}}$ \\
    
    UNI \cite{chen2024uni} & \underline{0.876} & \underline{0.816} & \textbf{0.736} & 0.675 & \underline{0.889} & 0.935 & 0.998 & 0.827 & 0.982 & 0.993 & 0.873$^{\textcolor{darkgreen}{0.010{{\uparrow}}}}$ \\
    
    H-Optimus-0 \cite{hoptimus0} & 0.872 & \textbf{0.836} & 0.697 & 0.685 & 0.790 & \textbf{0.944} & \textbf{1.000} & 0.881 & 0.971 & \textbf{0.996} & 0.867$^{\textcolor{darkgreen}{0.013{{\uparrow}}}}$ \\
    
    Virchow2 \cite{virchow2} & 0.831 & 0.735 & 0.693 &  \underline{0.732} & 0.863 & \underline{0.942} & 0.995 & 0.875 & 0.986 & \textbf{0.996} & 0.865$^{\textcolor{darkgreen}{0.023{{\uparrow}}}}$ \\
    
    \textbf{Phikon \cite{Filiot2023.07.21.23292757} (Ours)} & 0.803 & 0.780 & 0.699 & 0.685 & 0.851 & 0.938 & \textbf{1.000} & 0.830 & 0.977 & 0.986 & 0.855$^{\textcolor{darkgreen}{0.015{{\uparrow}}}}$ \\
    
    CTransPath \cite{ctranspath} & 0.800 & 0.788 & 0.678 & 0.723 & \textbf{0.895} & 0.923 & 0.896 & 0.838 & 0.977 & \textbf{0.996} & 0.851$^{\textcolor{darkgreen}{0.018{{\uparrow}}}}$ \\
    
    Kaiko (B/8) \cite{kaiko} & 0.788 & 0.734 & \underline{0.702} & 0.716 & 0.866 & 0.939 & 0.984 & 0.798 & 0.941 & \textbf{0.996} & 0.846$^{\textcolor{darkgreen}{0.017{{\uparrow}}}}$ \\
    
    Virchow \cite{virchow} & 0.801 & 0.800 & 0.681 & 0.718 & 0.788 & 0.940 & 0.989 & 0.793 & 0.966 & 0.987 & 0.846$^{\textcolor{darkgreen}{0.018{{\uparrow}}}}$ \\
    
    CONCH \cite{CONCH} & 0.835 & 0.782 & 0.700 & 0.604 & 0.846 & 0.922 & 0.981 & 0.830 & 0.941 & \underline{0.995} & 0.844$^{\textcolor{darkgreen}{0.020{{\uparrow}}}}$ \\
    
    Lunit-DINO (B/8) \cite{lunit_dino} & 0.836 & 0.749 & 0.648 & 0.649 & 0.822 & 0.927 & 0.998 & 0.818 & 0.971 & 0.994 & 0.841$^{\textcolor{darkgreen}{0.020{{\uparrow}}}}$ \\
    
    PathDINO-512 \cite{alfasly2023rotationagnostic} & 0.807 & 0.749 & 0.659 & 0.731 & 0.851 & 0.913 & 0.955 & 0.647 & 0.980 & 0.990 & 0.828$^{\textcolor{darkgreen}{0.012{{\uparrow}}}}$ \\
    
    Hibou (B/8) \cite{hibou} & 0.831 & 0.777 & 0.630 & 0.618 & 0.846 & 0.937 & 0.995 & 0.693 & 0.962 & 0.994 & 0.828$^{\textcolor{darkgreen}{0.021{{\uparrow}}}}$ \\
    
    Remedis \cite{azizirobust2023}& 0.716 & 0.731 & 0.578 & 0.715 & 0.711 & 0.911 & 0.958 & 0.760 & 0.965 & 0.994 & 0.804$^{\textcolor{darkgreen}{0.003{{\uparrow}}}}$ \\
    
    PLIP  \cite{huang2023visual} & 0.739 & 0.728 & 0.706 & 0.605 & 0.745 & 0.903 & 0.924 & 0.700 & 0.910 & 0.989 & 0.795$^{\textcolor{darkgreen}{0.021{{\uparrow}}}}$ \\
    
    DINOv2 ViT-L Imagenet \cite{oquab2024DINOv2} & 0.765 & 0.731 & 0.578 & 0.515 & 0.820 & 0.906 & 0.689 & 0.673 & 0.919 & 0.981 & 0.757$^{\textcolor{darkgreen}{0.051{{\uparrow}}}}$ \\
 
    \bottomrule
\end{tabular}
\vspace{0.5cm}
\caption{Table 2: Downstream performance (median AUC over 10,000 bootstrap runs) of feature extractors on eight different slide-level tasks. Evaluation protocol is detailed in \ref{evaluation-protocol} and evaluation tasks are detailed in \ref{evaluation-tasks}. For each task, best performance is in \textbf{bold}, second best performance is \underline{underlined}. \textbf{Average} column shows the mean performance across all tasks and models are ranked from best to last. $\textcolor{darkgreen}{0.051{{\uparrow}}}$ indicates that ensembling yields a 5.1 overall increase in AUC compared to a single-shot training with hyperparameters optimized during CV. $^{1}$ CTransPath was pre-trained on TCGA and Paip WSIs, hence performance on Paip cohort might be over-estimated.}
\label{table:downsteam-performance}
\end{table}

\begin{table}[ht!] 
\small
  \centering
  \addtolength{\leftskip} {-3.5cm}
  \addtolength{\rightskip}{-3.5cm}
\begin{tabular}{lccccc|ccccccccc}
 \toprule
 & GigaPath & H-0 & UNI & Phikon-v2 & Virchow2 & CTransPath & Phikon & Kaiko (B/8) & Virchow & CONCH & Lunit-DINO (B/8)  \\

GigaPath & \cellcolor{lightgray} & \cellcolor{lightgray} & \cellcolor{lightgray}  & \cellcolor{lightgreen}<0.001 & \cellcolor{lightorange} 0.001 & \cellcolor{lightgreen}<0.001 & \cellcolor{lightgreen}<0.001 & \cellcolor{lightgreen}<0.001 & \cellcolor{lightgreen}<0.001 & \cellcolor{lightgreen}<0.001 & \cellcolor{lightgreen}<0.001  \\

H-0 & \cellcolor{lightgray} & \cellcolor{lightgray} & \cellcolor{lightred} 0.031$^*$  & \cellcolor{lightgray}  & \cellcolor{lightred} 0.026 & \cellcolor{lightgreen}<0.001 & \cellcolor{lightgreen}<0.001 & \cellcolor{lightgreen}<0.001 & \cellcolor{lightgreen}<0.001 & \cellcolor{lightgreen}<0.001 & \cellcolor{lightgreen}<0.001  \\

UNI & \cellcolor{lightgray} & \cellcolor{lightgray} & \cellcolor{lightgray} &  \cellcolor{lightred} 0.020 & \cellcolor{lightgreen}<0.001 & \cellcolor{lightgreen}<0.001 & \cellcolor{lightgreen}<0.001 & \cellcolor{lightgreen}<0.001 & \cellcolor{lightgreen}<0.001 & \cellcolor{lightgreen}<0.001 & \cellcolor{lightgreen}<0.001 \\

Phikon-v2 & \cellcolor{lightgray} & \cellcolor{lightgray} & \cellcolor{lightgray} & \cellcolor{lightgray} & \cellcolor{lightgray} & \cellcolor{lightgreen}<0.001 & \cellcolor{lightgreen}<0.001 & \cellcolor{lightgreen}<0.001 & \cellcolor{lightgreen}<0.001 & \cellcolor{lightgreen}<0.001 & \cellcolor{lightgreen}<0.001  \\

Virchow2 & \cellcolor{lightgray} & \cellcolor{lightgray} & \cellcolor{lightgray} & \cellcolor{lightgray} & \cellcolor{lightgray} & \cellcolor{lightgreen}<0.001 & \cellcolor{lightred}0.016 & \cellcolor{lightgray} & \cellcolor{lightgreen}<0.001 & \cellcolor{lightgreen}<0.001 & \cellcolor{lightgreen}<0.001  \\
\hline

CTransPath & \cellcolor{lightgray} & \cellcolor{lightgray} & \cellcolor{lightgray} & \cellcolor{lightgray} & \cellcolor{lightgray} & \cellcolor{lightgray} & \cellcolor{lightgray} & \cellcolor{lightorange} 0.007 & \cellcolor{lightorange}  0.003 & \cellcolor{lightgray} & \cellcolor{lightorange} 0.006  \\

Phikon & \cellcolor{lightgray} & \cellcolor{lightgray} & \cellcolor{lightgray} & \cellcolor{lightgray} & \cellcolor{lightgray} & \cellcolor{lightgray} & \cellcolor{lightgray} & \cellcolor{lightgray} & \cellcolor{lightred} 0.046 & \cellcolor{lightgray} & \cellcolor{lightorange} 0.003  \\

Kaiko (B/8) & \cellcolor{lightgray} & \cellcolor{lightgray} & \cellcolor{lightgray} & \cellcolor{lightgray} & \cellcolor{lightgray} & \cellcolor{lightgray} & \cellcolor{lightgray} & \cellcolor{lightgray} & \cellcolor{lightgray} & \cellcolor{lightred} 0.043 & \cellcolor{lightorange} 0.001  \\

Virchow & \cellcolor{lightgray} & \cellcolor{lightgray} & \cellcolor{lightgray} & \cellcolor{lightgray} & \cellcolor{lightgray} & \cellcolor{lightgray} & \cellcolor{lightgray} & \cellcolor{lightgray} & \cellcolor{lightgray} &\cellcolor{lightgray} & \cellcolor{lightorange} 0.006 \\

\bottomrule
\end{tabular}
\vspace{0.5cm}
\caption{Table 3: Fisher-combined corrected p-values comparing pairwise models performance. A clear block composed of ViT-L+ models is standing out. Grey cells indicate no statistical significance at the 95\% confidence level. \textit{Reading note}: if $i$ and $j$ respectively denotes the ith row and jth column, cell $(i, j)$ indicates the p-value testing for model$_i$ statistical superiority over model$_j$. For instance, top left cell with $^*$ indicates that H-Optimus-0 shows higher performance over UNI overall (p=0.031).}
\label{table:pvalues}
\end{table}

Table \ref{table:downsteam-performance} shows the results of the slide-level evaluation described in section \ref{evaluation-protocol}. We compare Phikon-v2 against our previously released feature extractor (Phikon) and 14 other extractors from the literature (13 in-domain and one out-of-domain baseline). To identify general performance trends, it is necessary to aggregate performance across multiple tasks. A common practice consists in reporting mean or median performance over downstream tasks (see Table \ref{table:downsteam-performance} last column), yet statistical comparisons offer different insights on models \textit{general} behaviours and pairwise rankings. For this reason, we also highlight Fisher-combined corrected p-values in Table \ref{table:pvalues}, which indicates whether a given model $i$ \textit{at ith row} shows superior performance overall with respect to another model $j$ \textit{at jth column} from a statistical standpoint. This evaluation procedure less penalizes excellent models under-performing on a very limited number of tasks (e.g., H-Optimus-0 on IDH1, or UNI on MSI-Cy1), hence giving more reliable models rankings per se. We believe \ref{table:pvalues} should guide the way models are evaluated and compared in future benchmarks.

\subsection{ViT-L+ histology encoders stand out}

Tables \ref{table:downsteam-performance} and \ref{table:pvalues} consistently put GigaPath to the forefront. The later ranks first on 7 out of 10 downstream external cohorts. Similarly, H-Optimus-0 ranks first on 4 out of 7, explaining the absence of statistical difference between Bioptimus model and GigaPath in Table \ref{table:pvalues}. Phikon-v2 and UNI achieve excellent performance overall, respectively ranking first on 2 out of 7 tasks, and ranking second on 3 out of 7 tasks. Those two models share key components by design: they both leverage a ViT-L architecture pre-trained using DINOv2 on a dataset with same order of magnitude ($\sim$ 50k slides for Phikon-v2 with 400M tiles, $\sim$100k slides for UNI with 100M tiles). Note that both Phikon-v2 and UNI outperform Phikon with an overall gain in AUC of almost 2 points (p<1e-4). The later ranks sixth globally, being on-par with CTransPath (p=0.17), both models performing significantly better than Virchow (p=0.046 and p=0.003, respectively).

Our results highlight the pre-eminence of ViT-L+ models among our top 5. It should however be noted that performance (and rankings) of these feature extractors are relatively inconsistent across different evaluation tasks and datasets. For instance, GigaPath and UNI demonstrates strong performance on the Bcnb dataset, excelling in ER, PR, and HER2 validation tasks. However, their respective performance decline on the Herohe dataset, where Phikon-v2 emerges as the top performer followed by Virchow2. Conversely, Phikon-v2 excels in the MSI task but performs poorly on IDH1, similary to H-Optimus-0. This suggests that assessing the performance of a given feature extractor by looking solely at one or a few validation tasks is not enough to draw clear conclusions. A robust and fair evaluation requires several real-world datasets for each task as well as several candidate models, so as to evaluate the generalization capabilities of a given feature extractor and its sensitivity to distribution shifts, which matters most for clinical deployment. Moreover, a deeper understanding of how the pre-training samples influence downstream evaluation is needed. Indeed, Phikon performs better (+1.1\%, p=0.17) than Phikon-v2 on IDH1 task despite the absence of brain tissues during its pre-training; same applies for CTransPath performing best (0.894 [0.873 - 0.912]), on par with GigaPath which is 40x bigger, and surpassing H-Optimus-0 by a surprising +10.5 AUC margin.

\begin{figure}[ht!]
  \centering
  \includegraphics[width=\textwidth]{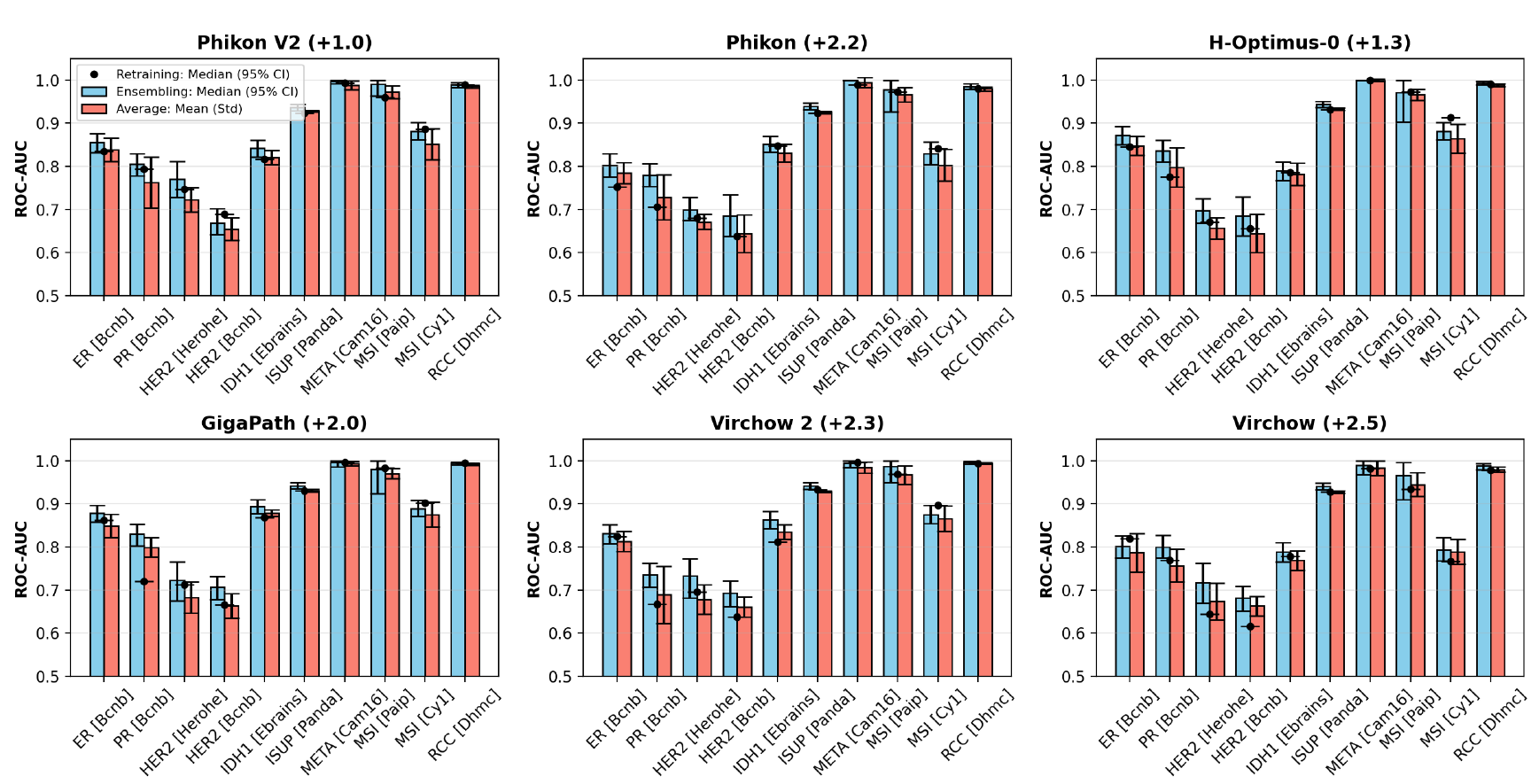}
  \caption{Figure 2: Comparison of ensembling performance against one-shot retraining. For each task, we compute the median AUC and 95\% confidence intervals across 10,000 repeats based on the ensembling of the 25 models' predictions issued from cross-validation (i.e., statistics over one predictions distribution, "Ensembling", blue). We compare this metric to the average (and standard deviation) of the 25 AUCs taken from each individual model without ensembling (i.e., statistics over 25 AUCs distribution, "Average", red). We eventually compare it to the one-shot retraining strategy, consisting in training a unique model on the whole training dataset for a number of epochs optimized during cross-validation ("Retraining", black dots). \textit{Reading note}: "(+1.0)" indicates that ensembling yields a +1.0 AUC gain over retraining, averaged across all tasks.}
  \label{ensembling}
\end{figure}

\begin{table}[ht!] 
  \centering
  \addtolength{\leftskip} {-3cm}
  \addtolength{\rightskip}{-3cm}
  \begin{tabular}{llll|l}
    \toprule
    \multicolumn{1}{c}{} & 
    \multicolumn{3}{c}{MSI} \\
    \cmidrule(r){2-4}
    \textbf{Extractor} & Cy1 & Ngx1 & Paip & \textbf{Average} \\
    \midrule
    \textbf{iBOT (B/16) Coad (Ours) \cite{Filiot2023.07.21.23292757}} & \textbf{0.899}&	0.945&	\underline{0.989}	&\textbf{0.944} \\
    GigaPath \cite{gigapath} & \underline{0.888}&	\underline{0.949}	&0.980&	\underline{0.939}$^{ n.s.}$  \\
     H-Optimus-0 \cite{hoptimus0} &  0.881	&\textbf{0.961}&	0.971	&0.938$^{ n.s.}$\\
     Virchow2 \cite{virchow2} & 0.875	&0.938	&0.986&	0.933$^*$ \\
    Phikon-v2 (Ours) & 0.882&	0.919&	\textbf{0.991}	&0.931$^*$ \\
    UNI \cite{chen2024uni} & 0.827&	0.926&	0.982	&0.911$^{***}$ \\
    Phikon \cite{Filiot2023.07.21.23292757} (Ours) & 0.830&	0.921	&0.977&	0.909$^{***}$ \\
    Virchow \cite{virchow} & 0.793&	0.879	&0.966&	0.879$^{***}$ \\
    \bottomrule
\end{tabular}
\vspace{0.5cm}
\caption{Table 4: Downstream performance (median AUC over 10,000 bootstrap runs) of feature extractors on MSI prediction tasks. Evaluation protocol is detailed in \ref{evaluation-protocol} and evaluation tasks are detailed in \ref{evaluation-tasks}. For each task, best performance is in \textbf{bold}, second best performance is \underline{underlined}. \textbf{Average} column shows the mean performance across all tasks and models are ranked from best to last. \textbf{iBOT (B/16) Coad (Ours)} is a ViT-B pre-trained on 4 million tiles from TCGA-COAD. \textit{Reading note}: we provide p-values for comparing our specialized model to others following the same procedure as for Table \ref{table:pvalues}: $^{ n.s.}:$ non-significant, $^{*}: p<0.05$, $^{**}: p<0.01$, $^{***}: p<0.001$.}
\label{table:downsteam-performance-msi}
\end{table}

\subsection{Don't set aside specialized models}

Table \ref{table:downsteam-performance-msi} compares \textbf{iBOT (B/16) Coad (Ours)} against top foundation models on MSI prediction task in colorectal cancer. The former is a ViT-B pre-trained on 4 million tiles from TCGA-COAD cohort for 165k iterations using iBOT \cite{Filiot2023.07.21.23292757}. This model, 13x smaller and pre-trained on 350x less histology images than Virchow2, ranks first on our benchmark made of 3 external validation cohorts (2 private, 1 public being Paip). Obviously, these results suggest that scaling per se is not - as of now, the systematic solution for solving biomarker prediction tasks. This is absolutely critical for the validation, clinical deployment and adoption of diagnostic and pre-screening tools, may it be working on MSI prediction \cite{saillard_blind_2023} or any other biomarker \cite{wang2024screenallhighthroughputpancancer}. Note that pre-training with DINOv2 slightly decreased performance compared to iBOT, which may suggest the DINOv2 superiority over iBOT (both methods beings conceptually very similar) is not straightforward for lighter models.

\subsection{Ensembling yields consistent improvements over one-shot retraining}

Lastly as shown on Figure \ref{ensembling}, our ensembling strategy  designed for slide-level downstream tasks yields strong improvements over one-shot retraining (see Section \ref{evaluation-protocol} for details), a very popular yet sub-optimal practice. Indeed, ensembling yields a statistically significant AUC increase over one-shot retraining: first for each histology feature extractor taken individually (p<0.0001, except for Lunit-DINO (B/8) with p=0.07 and REMEDIS with p=0.47), and overall when testing across all feature extractors and tasks with a +1.75 AUC increase (p<0.001). This shows the remarkable benefit of ensembling at a minor extra computation cost. This strategy allows for more stable and robust predictions\footnote{See \url{https://www.kaggle.com/competitions/UBC-OCEAN/discussion/466455}.} and should be favored in forthcoming benchmarks specific to weakly-supervised tasks.

\section{Conclusion}

Phikon-v2 significantly improves upon Phikon by employing a new self-supervised approach, a wider architecture, and a larger collection of public datasets for pre-training. It demonstrates competitive performance compared to current state-of-the-art methods across a range of 8 slide-level biomarker prediction tasks covering 5 external datasets unseen during pre-training. As with Phikon, we are publicly releasing Phikon-v2 under a non-commercial license, enabling widespread access, collaboration, and further research and development in the field.

In light of recent literature, our scaling experiments with DINOv2 have been successful and marked an improvement over iBOT which was used to pre-train our previous foundation model. Nonetheless, we did not assess the scalability of iBOT-only on PANCAN-XL, which is led to further research. Other Masked Image Modelling methods may also bring improvements over DINOv2 or iBOT \cite{haghighat2024pretraining}.

Our evaluation of a large and exhaustive panel of feature extractors\footnote{We plan to update our benchmarks with Hibou-Large \cite{hibou} and GPFM \cite{ma2024generalizablepathologyfoundationmodel}.} proposed by the community has also confirmed the undeniable benefit of in-domain self-supervised pre-training over out-of-domain pre-training, even for models that are 30 times smaller \cite{alfasly2023rotationagnostic}. 

While GigaPath, H-Optimus-0, Phikon-v2, UNI and Virchow2 all reinforce the benefits of FM scaling to improve downstream performance, these models remain general-purpose by design in terms of downstream tasks and target organs. Consequently, with the extensive range of models available in the community, it is likely that some may be better suited to specific tasks and organs, a trend already observed in other benchmarks \cite{campanella2024clinicalbenchmarkpublicselfsupervised} and highlighted in this work specifically for MSI prediction. Ensembling at the feature extractor is an existing avenue \cite{neidlinger2024benchmarkingfoundationmodelsfeature} yet comes with a significant computational bottleneck. We believe that, similar to the specialization seen in large language models, an important focus should be placed on organ- and task-specific fine-tuning \cite{hu2021lora} or distillation \cite{marrie2024good} of foundational models for clinical deployment \cite{nicke2024tissueconceptssupervisedfoundation}. This focus is crucial as scaling experiments reach an inflexion point where significant improvements become challenging without access to massive datasets and substantial computing power. 

Future efforts should continue focusing on developing self-supervised learning methods intrinsically tailored to histology data, either with scaling \cite{prov_gigapath, virchow2} or without \cite{ctranspath,alfasly2023rotationagnostic,jaume2024multistainpretrainingsliderepresentation}. As of now, too little attention has been given to assessing and predicting the robustness of histology foundation models to slide preparation and acquisition shifts \cite{wolflein_good_2023}, which is crucial for bridging the gap between development and clinical use. This is why evaluating FMs requires real-world and open evaluations relevant to clinical applications, encompassing specific downstream tasks over various external validations and covering a wide variety of potential domain shifts and population bias \cite{fairness_mahmood}.

\section*{Acknowledgments}

We thank DINOv2 authors for their research, repository as well as their responsive feedbacks.

\textit{Computing resources}

This work was granted access to the High-Performance
Computing (HPC) resources of IDRIS under the allocation 2023-A0141012519 made by
GENCI.

\textit{Data sources}

The results published here are partly based upon data
generated by the TCGA Research Network:
https://www.cancer.gov/tcga. Regarding the
PAIP dataset, de-identified pathology images and
annotations used in this research were prepared and
provided by the Seoul National University Hospital by a
grant of the Korea Health Technology R\&D Project through
the Korea Health Industry Development Institute (KHIDI),
funded by the Ministry of Health \& Welfare, Republic of
Korea (grant number: HI18C0316). The Genotype-Tissue Expression (GTEx) Project was supported by the Common Fund of the Office of the Director of the National Institutes of Health, and by NCI, NHGRI, NHLBI, NIDA, NIMH, and NINDS. The data used for the analyses described in this manuscript were obtained from the GTEx Portal on 07/01/2023.

\small
\bibliographystyle{unsrt}
\bibliography{biblio}

\newpage

\section*{Supplementary data}

\begin{table}[ht!]
\centering
\begin{tabular}{l|l}
\hline Hyper-parameter & Value \\
\hline Layers & 24 \\
Heads & 16 \\
Patch size & 16 \\
FFN layer & MLP \\
Head activation & GELU \\
Embedding dimension & 1024 \\
Stochastic dropout rate & 0.3 \\
Layerscale & 1e-5 \\
\hline Global crop scale & (0.32, 1.0) \\
Global crop number (size) & 2 (224) \\
Local crop scale & (0.05, 0.32) \\
Local crop number (size) & 8 (96) \\
Max masking ratio & 0.5 \\
Min masking ratio & 0.1 \\
iBOT mask sample probability & 0.5 \\
Gradient clipping max norm & 3.0 \\
Normalize last layer & $\checkmark$ \\
Shared head & x \\
Registers & x \\
\hline Adam $\beta$ & (0.9,0.999) \\
Batch size & 4096 \\
Freeze last layer iterations & 1250 \\
Warmup iterations & 25000 \\
Warmup teacher temperature iterations & 75000 \\
High-resolution finetuning iterations & 0 \\
Max Iterations & 250000 \\
Learning rate schedule & Cosine \\
Learning rate (start) & 0 \\
Learning rate (post warmup) & 4e-3 \\
Learning rate (final) & 1e-6 \\
Teacher temperature (start) & 0.04 \\
Teacher temperature (final) & 0.4 \\
Teacher momentum (start) & 0.992 \\
Teacher momentum (final) & 1.000 \\
Weight decay (start) & 0.04 \\
Weight decay (end) & 0.4 \\
Automatic mixed precision & FP16 \\
\hline
\end{tabular}
\vspace{0.5cm}
\caption{Extended Table 5: DINOv2 hyperparameters used for Phikon-v2 pretraining. 32 nodes with 4 × 32GB NVIDIA V100
GPUs were used for training.}
\label{table:hyperparams}
\end{table}

\begin{center}
\begin{tabular}{llr}
\toprule
Organ & Dataset & No. WSIs \\
\midrule
\multirow{2}{*}{Adipose} & GTEx Adipose Visceral Omentum & 538 \\
                         & GTEx Adipose Subcutaneous & 656 \\
\midrule
\multirow{3}{*}{Adrenal Gland} & TCGA-ACC (FF) & 95 \\
                                & GTEx Adrenal Gland & 212 \\
                                & TCGA-ACC (FFPE) & 227 \\
\midrule
\multirow{3}{*}{Artery} & GTEx Artery Coronary & 239 \\
                        & GTEx Artery Aorta & 430 \\
                        & GTEx Artery Tibial & 656 \\
\midrule
\multirow{3}{*}{Bladder} & GTEx Bladder & 21 \\
                         & TCGA-BLCA (FFPE) & 455 \\
                         & TCGA-BLCA (FF) & 469 \\
\midrule
\multirow{3}{*}{Blood-related Diseases} & TCGA-DLBC (FFPE) & 44 \\
                                        & TCGA-DLBC (FF) & 59 \\
                                        & CPTAC-AML & 122 \\
\midrule
\multirow{7}{*}{Brain} & UPENN-GBM & 67 \\
                       & GTEx Brain Cerebellum & 240 \\
                       & GTEx Brain Cortex & 254 \\
                       & CPTAC-GBM & 462 \\
                       & TCGA-LGG (FF) & 728 \\
                       & TCGA-LGG (FFPE) & 844 \\
                       & TCGA-GBM (FFPE) & 859 \\
                       & TCGA-GBM (FF) & 1192 \\
\midrule
\multirow{9}{*}{Breast} & Post-NAT-BRCA & 96 \\
                        & Breast Metastases (MSKCC) & 130 \\
                        & CyRx (private) & 199 \\
                        & HER2 Tumor ROIs (v3) & 276 \\
                        & GTEx Breast Mammary Tissue & 457 \\
                        & CPTAC-BRCA & 653 \\
                        & TCGA-BRCA (FFPE) & 1118 \\
                        & TCGA-BRCA (FF) & 1977 \\
                        & TUH DPath Breast & 3504 \\
\midrule
\multirow{14}{*}{Colorectal} & Biobank-CMB-CRC (v1) & 31 \\
                             & TCGA-READ (FFPE) & 128 \\
                             & Ngx (private) & 199 \\
                             & Hungarian Colorectal Screening (v1 update) & 200 \\
                             & TCGA-READ (FF) & 302 \\
                             & GTEx Colon Sigmoid & 370 \\
                             & TCGA-COAD (FFPE) & 371 \\
                             & CPTAC-COAD & 372 \\
                             & GTEx Colon Transverse & 405 \\
                             & TCGA-COAD (FF) & 840 \\
                             & Mpth Rx Philips (private) & 998 \\
                             & Mpth Rx Roche (private) & 1001 \\
\midrule
Eye & TCGA-UVM (FF) & 70 \\
\midrule
\multirow{7}{*}{Gastroesophageal} & Biobank-CMB-GEC (v1) & 3 \\
                                  & TCGA-ESCA (FFPE) & 158 \\
                                  & TCGA-ESCA (FF) & 238 \\
                                  & GTEx Stomach & 356 \\
                                  & TCGA-STAD (FFPE) & 400 \\
                                  & GTEx Esophagus Muscularis & 512 \\
                                  & GTEx Esophagus Mucosa & 550 \\
                                  & TCGA-STAD (FF) & 755 \\
\midrule
\multirow{3}{*}{Head and Neck} & CPTAC-HNSCC & 390 \\
                               & TCGA-HNSC (FFPE) & 472 \\
                               & TCGA-HNSC (FF) & 727 \\
\midrule
\multirow{2}{*}{Heart} & GTEx Heart Atrial Appendage & 427 \\
                       & GTEx Heart Left Ventricle & 429 \\
\bottomrule
\end{tabular}
\end{center}

\begin{center}
\begin{tabular}{llr}
\toprule
Organ & Dataset & No. WSIs \\
\midrule
\multirow{8}{*}{Kidney} & GTEx Kidney Medulla & 4 \\
                        & GTEx Kidney Cortex & 85 \\
                        & TCGA-KICH (FFPE) & 121 \\
                        & TCGA-KICH (FF) & 205 \\
                        & TCGA-KIRP (FFPE) & 284 \\
                        & TCGA-KIRP (FF) & 473 \\
                        & TCGA-KIRC (FFPE) & 519 \\
                        & CPTAC-CCRCC & 783 \\
                        & TCGA-KIRC (FF) & 1654 \\
\midrule
\multirow{4}{*}{Liver} & TCGA-CHOL (FFPE) & 38 \\
                       & GTEx Liver & 224 \\
                       & TCGA-LIHC (FFPE) & 372 \\
                       & TCGA-LIHC (FF) & 491 \\
\midrule
\multirow{8}{*}{Lung} & Biobank-CMB-LCA (v1) & 13 \\
                      & PennyCuick & 112 \\
                      & TCGA-LUSC (FFPE) & 512 \\
                      & TCGA-LUAD (FFPE) & 531 \\
                      & GTEx Lung & 576 \\
                      & TCGA-LUAD (FF) & 1067 \\
                      & CPTAC-LSCC & 1081 \\
                      & TCGA-LUSC (FF) & 1100 \\
                      & NLST-pathology-1225 & 1225 \\
\midrule
\multirow{2}{*}{Mesothelium} & TCGA-MESO (FFPE) & 87 \\
                             & TCGA-MESO (FF) & 88 \\
\midrule
Minor Salivary Gland & GTEx Minor Salivary Gland & 162 \\
\midrule
\multirow{5}{*}{Muscle - Skeletal} & Biobank-CMB-MML (v1) & 2 \\
                                   & CPTAC-SAR & 273 \\
                                   & TCGA-SARC (FF) & 290 \\
                                   & TCGA-SARC (FFPE) & 600 \\
                                   & GTEx Muscle Skeletal & 797 \\
\midrule
\multirow{3}{*}{Others} & TCGA-PCPG (FF) & 189 \\
                        & TCGA-PCPG (FFPE) & 195 \\
                        & GTEx Nerve Tibial & 615 \\
\midrule
\multirow{6}{*}{Ovary} & TCGA-OV (FFPE) & 107 \\
                       & GTEx Ovary & 179 \\
                       & CPTAC-OV & 221 \\
                       & Ovarian Bevacizumab Response & 286 \\
                       & PTRC-HGSOC & 348 \\
                       & TCGA-OV (FF) & 1373 \\
\midrule
\multirow{4}{*}{Pancreas} & TCGA-PAAD (FFPE) & 209 \\
                          & TCGA-PAAD (FF) & 257 \\
                          & GTEx Pancreas & 327 \\
                          & CPTAC-PDA & 557 \\
\midrule
Pituitary Gland & GTEx Pituitary & 282 \\
\midrule
\multirow{5}{*}{Prostate} & Biobank-CMB-PCA (v1) & 3 \\
                          & GTEx Prostate & 242 \\
                          & TCGA-PRAD (FFPE) & 449 \\
                          & NADT Prostate & 488 \\
                          & TCGA-PRAD (FF) & 723 \\
\midrule
\multirow{8}{*}{Skin} & Biobank-CMB-MEL (v1) & 33 \\
                      & TCGA-UVM (FFPE) & 80 \\
                      & Hodis & 148 \\
                      & CPTAC-CM & 405 \\
                      & TCGA-SKCM (FFPE) & 475 \\
                      & TCGA-SKCM (FF) & 475 \\
                      & GTEx Skin Not Sun Exposed Suprapubic & 601 \\
                      & GTEx Skin Sun Exposed Lower Leg & 696 \\        
\midrule
Small Intestine & GTEx Small Intestine Terminal Ileum & 186 \\
\bottomrule
\end{tabular}
\end{center}

\begin{table}
\begin{center}
\begin{tabular}{llr}
\toprule
Organ & Dataset & No. WSIs \\
\midrule
Spleen & GTEx Spleen & 241 \\
\midrule
\multirow{3}{*}{Testis} & TCGA-TGCT (FF) & 156 \\
                        & TCGA-TGCT (FFPE) & 254 \\
                        & GTEx Testis & 359 \\
\midrule
\multirow{2}{*}{Thymus} & TCGA-THYM (FF) & 137 \\
                        & TCGA-THYM (FFPE) & 180 \\
\midrule
\multirow{3}{*}{Thyroid} & TCGA-THCA (FFPE) & 519 \\
                         & TCGA-THCA (FF) & 639 \\
                         & GTEx Thyroid & 650 \\
\midrule
\multirow{12}{*}{Uterus} & GTEx Cervix Ectocervix & 9 \\
                         & GTEx Fallopian Tube & 9 \\
                         & GTEx Cervix Endocervix & 10 \\
                         & TCGA-UCS (FF) & 63 \\
                         & TCGA-UCS (FFPE) & 87 \\
                         & GTEx Uterus & 141 \\
                         & GTEx Vagina & 155 \\
                         & TCGA-CESC (FFPE) & 279 \\
                         & TCGA-CESC (FF) & 325 \\
                         & TCGA-UCEC (FFPE) & 566 \\
                         & TCGA-UCEC (FF) & 805 \\
                         & CPTAC-UCEC & 874 \\
\bottomrule
\end{tabular}
\vspace{0.5cm}
\caption{Extended Table 6: description of PANCAN-XL pre-training dataset. The total number of WSIs is 58,359. Unless specified otherwise, all datasets are publicly available.}
\label{table:pancanxl_details}
\end{center}
\end{table}

\end{document}